\documentclass[11pt,a4paper,reqno]{amsart}
\usepackage{amsfonts}
\usepackage{graphicx}   
\usepackage{graphics}
\usepackage{epsfig}

\usepackage{anysize}
\marginsize{3cm}{3cm}{3cm}{3cm}

\newtheorem{theorem}{Theorem}[section]

\makeatletter

\begin{document}

\setcounter{page}{1}
\bigskip
\bigskip

\title[\centerline{B.A.Rajabov: The theory of representations of groups $SO_{0}(2,1)$ and $ISO(2,1)$.
\hspace{0.5cm}}] {The theory of representations of groups $SO_{0}(2,1)$\\and $ISO(2,1)$. WIGNER COEFFICIENTS OF THE GROUP $SO_{0}(2,1)$}

\author[\hspace{0.7cm}\centerline{B.A.Rajabov: The theory of representations of groups $SO_{0}(2,1)$ and $ISO(2,1)$}] {B.A. RAJABOV$^1$}

\thanks {\noindent $^1$ N.Tusi Shamakhi Astrophysics Observatory, National Academy of Sciences of Azerbaijan\\
\indent \,\,\, e-mail: balaali.rajabov@mail.ru;\\}


\bigskip
\begin{abstract}
This paper is devoted to the representations of the groups $SO (2,1)$ and $ISO (2,1)$. Those groups have an important role in cosmology, elementary particle theory and mathematical physics. Irreducible unitary representations of the principal continuous and supplementary as well as discrete series were obtained. Explicit expressions for spherical functions of the group $SO_0(2,1)$ are obtained through the Gauss hypergeometric functions. The Wigner coefficients of the group $SO_{0}(2,1)$ were computed and their explicit expressions using the bilateral series were represented. The results could be used to study the non-degenerate representations of the de Sitter group $SO(3,2)$.


\end{abstract}

\maketitle
\bigskip


\section{Introduction}
Representations of the group  $SO_{0}(2,1)$ appear in various problems of theoretical and mathematical physics, in particular, as a group of motions in general theory of relativity and dual models of hadron reactions [1-4]. Representations of groups $SU(1,1)$ and $SL(2,R)$, which are universal covering groups for the group $SO_{0}(2,1)$ and are locally isomorphic to it, were studied in [5-7], using methods of integral geometry and invariant bilinear forms.

On the other hand, the group $SO_{0}(2,1)$ is the normal divisor of a non-homogeneous group $ISO(2,1)$ - the group of motions in the three-dimensional pseudo-Euclidean space. The group $ISO(2,1)$ is the stationary subgroup of a surface of transitivity, namely the cone of the de Sitter group $SO(3,2)$. Because of  $ISO(2,1)$ and its representations play an essential role in the study of non-degenerate representations of the de Sitter group $SO(3,2)$.

This article is an introductory part of the work devoted to the theory of non-degenerate representations and Wigner coefficients of the de Sitter group $SO(3,2)$. Since it is of independent interest, we decided to publish it separately.

\section{IRREDUCIBLE UNITARY REPRESENTATIONS OF $SO_{0}(2,1)$}
The group  $SO_{0}(2,1)$ is a connected component unit of the group of motions in the 3-dimensional pseudo-Euclidean space retaining invariant the following quadratic form:
\[
\left[k,k\right]=k_{0}^{2}-k_{1}^{2}-k_{2}^{2}.
\]

It is three-parametric, similar to the group  of rotations of 3-dimensional Euclidean space. We introduce the following system of coordinates on the upper field of a hyperboloid $[k,k]=1,\;k_{0}>0$:
\[
[k,k]=(\cosh\alpha,\sinh\alpha\sin\phi,\sinh\alpha\cos\phi).
\]

Since the upper field of the hyperboloid $[k,k]=1,\;k_{0}>0$, is a transitive surface with $k=(1,0,0)$ as a fixed point one can get the following decomposition:
\begin{equation}
g=r(\phi_{1})h(\alpha)r(\phi_{2}),
\end{equation}
where $r(\cdot)$  is a rotation on the plane $(k_{1},k_{2})$ and $h(\cdot)$ is a hyperbolic rotation on the plane $(k_{0},k_{2})$.

The representation of the group $SO_{0}(2,1)$ will be constructed in the space of infinitely differentiable homogeneous functions $\mathcal{F^{\sigma}}$, defined on the upper field of a cone without the vertex $[k,k]=1,\;k_{0}>0$, and the degree of homogeneity $\sigma$:
\begin{equation}
F(ak)=a^{\mathit{\sigma}}F(k),\qquad a>0,
\end{equation}
\begin{equation}
T(g)F(k)=F\left(g^{-1}k\right),\qquad g\in SO(2,1).
\end{equation}

Introduce the following system of coordinates on the cone  $[k,k]=0,\;k_{0}>0$:
\begin{equation}
k=\omega(1,\sin\phi,\cos\phi),\quad\omega>0,\quad0\leq\phi<2\pi.
\end{equation}

Then applying (2)-(4) one can establish an isomorphism between the space $\mathcal{F^{\sigma}}$ and the space of infinitely differentiable functions on the circle {$S^{1}$:
\begin{equation}
F(k)=\omega^{\sigma}f(\phi),\qquad f(\phi)=F(k)|_{\omega=1}.
\end{equation}

We use the same notation $\mathcal{F^{\sigma}}$ for the space of infinitely differentiable functions on the circle. The usual Fourier decomposition forms a canonical basis:
\begin{equation}
f(\phi)=\sum_{m=-\infty}^{+\infty}\,f_{m}e^{im\phi}.
\end{equation}
It is obvious that a restriction of the representation (3) to the subgroup of rotations on the plane $(k_{1},k_{2})$} is an additive group on the circle:
\begin{equation}
T\left(r(\phi_{1})\right)f(\phi)=f(\phi-\phi_{1}).
\end{equation}

The following formula gives the restriction of the representation (3) to the subgroup of hyperbolic rotations on the plane $(k_{0},k_{2})$\footnote{The representations of the group $SO_0(2,1)$ in contrast to the representations of its universal covering groups $SU (1,1)$ and $SL(2,R)$ do not have a parity and corresponds to even representations of these groups. To construct representations of the group $SO_0(2,1)$ with a certain parity, it suffices to replace the circle  $\left\{ S^{1}:\,0\leq\phi<2\pi\right\} $ by its double cover $\left\{ S^{1}:\,0\leq\phi<4\pi\right\} $.}:
\begin{equation}
T\left(h(\alpha)\right)f(\phi)=\omega_{\alpha}^{\sigma}f\left(\phi_{\alpha}\right),
\end{equation}
where
\begin{eqnarray}
\omega_{\alpha} & = & \cosh\alpha-\sinh\alpha\cos\phi,\nonumber \\
\sin\phi_{\alpha} & = & \sin\phi/\omega_{\alpha},\nonumber \\
\cos\phi_{\alpha} & = & \left(-\sinh\alpha+\cosh\alpha\cos\phi\right)/\omega_{\alpha}.
\end{eqnarray}

Irreducible unitary representations of continuous principal series are obtained by introducing the following scalar product in \textbf{$\mathcal{F^{\sigma}}$}:\footnote{Here the sign “bar” denotes the complex conjugate. }
\begin{equation}
\left(f^{(1)},f^{(2)}\right)=\frac{1}{2\pi}\intop_{0}^{2\pi}\,\overline{f^{(1)}(\phi)}f^{(2)}(\phi)d\phi,
\end{equation}

It follows from (9) that the measure $d\phi$ at the hyperbolic rotations (8) is transformed as:
\begin{equation}
d\phi_{\alpha}=d\phi/\omega_{\alpha}.
\end{equation}

Obviously, it is invariant under rotations (7). It is also apparent that functions $e^{im\phi}$ form the orthonormal canonical basis concerning the scalar product (10). It follows from (7)-(9) and (11) that the scalar product (10) is invariant if the degree of homogeneity $\sigma$  is as follows: 
\begin{equation}
\sigma=-1/2+i\rho,\qquad-\infty<\varrho<+\infty,
\end{equation}

Unitarity and irreducibility of other series of representations in this space are studied by constructing invariant Hermit-bilinear functionals:
\begin{equation}
(f^{(1)},f^{(2)})=\frac{1}{4\pi^{2}}\intop_{0}^{2\pi}\intop_{0}^{2\pi}K(\phi_{1},\phi_{2})\overline{f^{(1)}(\phi_{1})}f^{(2)}(\phi_{2})d\phi_{1}d\phi_{2}.
\end{equation}
Here $K(\phi_{1},\phi_{2})$ is a generalized function over the space $\mathcal{F}^{\sigma}\otimes\mathcal{F}^{\sigma}$, [8]. 

The invariance condition and also formulas (7)-(8) and (11) imply that $K(\phi_{1},\phi_{2})$ is a function, which:
\begin{itemize}
	\item $K(\phi_{1},\phi_{2})$ -- depends on the difference of variables;
	\item $K(\phi_{1},\phi_{2})$ -- satisfies the following equation:
	\begin{equation}
	K(\phi_{1\alpha},\phi_{2\alpha})=\omega_{\alpha}^{\bar{\sigma}+1}\left(\phi_{1}\right)\omega_{\alpha}^{\sigma+1}\left(\phi_{2}\right)K(\phi_{1},\phi_{2}),
	\end{equation}		
\end{itemize}
where $\omega_{\alpha}(\cdot)$ is defined in (9). One can show that the solution to the equation (14) exists only when $\bar{\sigma}=\sigma,$ and this solution is:
\begin{equation}
K(\phi_{1},\phi_{2})=c\left[1-\cos\left(\phi_{1}-\phi_{2}\right)\right]^{-\sigma-1}.
\end{equation}
Here $c$ -- is an arbitrary constant.

Thus the invariant Hermit-bilinear form in the space $\mathcal{F}^{\sigma}\otimes\mathcal{F}^{\sigma}$ can be expressed as:
\begin{equation}
(f^{(1)},f^{(2)})=\frac{c}{4\pi^{2}}\intop_{0}^{2\pi}\intop_{0}^{2\pi}\left[1-\cos\left(\phi_{1}-\phi_{2}\right)\right]^{-\sigma-1}\overline{f^{(1)}(\phi_{1})}f^{(2)}(\phi_{2})d\phi_{1}d\phi_{2},
\end{equation}
where {$\sigma$} -- is a real number.

In order to find the canonical expression (16) and to study non-degeneracy and irreducibility of representations, it is necessary to find the Fourier expansion for the interior of this integral representation. Since this expansion will be used to study equivalence of representations of Wigner coefficients we consider complex values of $\sigma$, (see Appendix).

\section{COMPLEMENTARY AND DISCRETE SERIES OF REPRESENTATIONS}
It follows from (68) that in order to provide regularization of the form (16) it suffices to choose the arbitrary constant $c$ as: 
\[
c=c_{0}/\Gamma(-\sigma-1/2), 
\]
since for this form $\lambda=-\sigma-1$. Here $c_{0}$ is an arbitrary constant. 

In conclusion, for the invariant Hermit-bilinear form in the space $\mathcal{F}^{\sigma}\otimes\mathcal{F}^{\sigma}$ we have the following integral representation:
\begin{equation}
(f^{(1)},f^{(2)})=\frac{c_{0}}{4\pi^{2}\Gamma(-\sigma-1/2)}\intop_{0}^{2\pi}\intop_{0}^{2\pi}\left[1-\cos\left(\phi_{1}-\phi_{2}\right)\right]^{-\sigma-1}\overline{f^{(1)}(\phi_{1})}f^{(2)}(\phi_{2})d\phi_{1}d\phi_{2}.
\end{equation}

Recall that here $\sigma$ is a real number.

Applying (6) and (68) one can get the canonical form of the functional (17):
\begin{equation}
(f^{(1)},f^{(2)})=c_{0}\,\frac{2^{-\sigma-1}}{\sqrt{\pi}\Gamma(\sigma+1)}\sum_{m=-\infty}^{+\infty}\Phi_{m}\overline{f_{m}^{(1)}}f_{m}^{(2)},
\end{equation}
here 
\begin{equation}
\Phi_{m}=\frac{\Gamma(m+\sigma+1)}{\Gamma(m-\sigma)}.
\end{equation}

It follows from (19) that the form (18) to be positive definite it is necessary and sufficient that the following condition is satisfied:
\[
\frac{m+\sigma+1}{m-\sigma}>0
\]
for all integer $m$ and real numbers $\sigma$. This condition is satisfied only when $-1<\sigma<0$. It is obvious that the form (18)-(19) is non-degenerate for
non-integer $\sigma$. Thus we get:
\begin{theorem}
	The representation (7)-(8) in the space $\mathcal{F^{\mathit{\sigma}}}$ is irreducible unitary with respect to the scalar product (18) for $-1<\sigma<0$. This representation is called the complementary series.
\end{theorem}

For integer values of the $\sigma$ the form (18)-(19), obviously, degenerates.
Selecting suitably the arbitrary constant $c_{0}$, we can extend the bilinear form (17) on integer values $\sigma$. As a result we have the following degeneracy subspaces for integer values $\sigma$.

Form (18) degenerates on subspaces  $\mathcal{F_{\text{+}}^{\mathit{\sigma}}}$ and $\mathcal{F_{\text{-}}^{\mathit{\sigma}}}$, where
\begin{eqnarray*}
	\mathcal{F_{\text{+}}^{\mathit{\sigma}}} & = & \left\{ \Phi_{m}=0,\quad m=-\sigma,\,-\sigma+1,\,-\sigma+2,\ldots,\,+\infty\right\} .\\
	\mathcal{F_{\text{-}}^{\mathit{\sigma}}} & = & \left\{ \Phi_{m}=0,\quad m=\sigma,\,\sigma-1,\,\sigma-2,\ldots,\,-\infty\right\} 
\end{eqnarray*}

The space $\mathcal{F_{\text{+}}^{\mathit{\sigma}}}$ consists of functions of the form:
\[
f(\phi)=\sum_{m=-\sigma}^{+\infty}\,f_{m}e^{im\phi}=e^{i\sigma\phi}\sum_{m=0}^{+\infty}\,f_{m}^{'}e^{im\phi}.
\]

The space $\mathcal{F_{\text{-}}^{\mathit{\sigma}}}$ consists of functions of the form:
\[
f(\phi)=\sum_{m=-\infty}^{\sigma}\,f_{m}e^{im\phi}=e^{i\sigma\phi}\sum_{m=0}^{+\infty}\,f_{m}^{'}e^{-im\phi}.
\]

It is clear that the union of these subspaces $\mathcal{F_{\text{+}}^{\mathit{\sigma}}}\bigcup\mathcal{F_{\text{-}}^{\mathit{\sigma}}}$ is also a subspace of degeneration.

On non-negative integer values of $\sigma$ these subspaces intersect:
\[
\mathcal{F_{\text{0}}^{\mathit{\sigma}}}=\mathcal{F_{\text{+}}^{\mathit{\sigma}}}\bigcap\mathcal{F_{\text{-}}^{\mathit{\sigma}}},\qquad\sigma=0,1,2,\ldots
\]

It is obvious that the form (19) degenerates on a subspace $\mathcal{F_{\text{0}}^{\mathit{\sigma}}}$, where
\begin{eqnarray*}
	\mathcal{F_{\text{0}}^{\mathit{\sigma}}} & = & \left\{ \Phi_{m}=0,\quad m=-\sigma,\:-\sigma+1,\ldots,\,\,\sigma-1,\,\sigma\right\} .
\end{eqnarray*}

The space $\mathcal{F_{\text{0}}^{\mathit{\sigma}}}$ consists of functions of the form:
\[
f(\phi)=\sum_{m=-\sigma}^{\sigma}\,f_{m}e^{im\phi}
\]

Thus, we get the following result:
\begin{theorem}
For integral values of $\sigma$ the representation (7)-(8) of the group $SO(2,1)$ acting in the factor-spaces $\mathcal{F^{\mathit{\sigma}}}/\mathcal{F_{\text{+}}^{\mathit{\sigma}}}$,
$\mathcal{F^{\mathit{\sigma}}}/\mathcal{F_{\text{-}}^{\mathit{\sigma}}}$ and $\mathcal{F^{\mathit{\sigma}}}/\mathcal{F_{\text{+}}^{\mathit{\sigma}}}\bigcup\mathcal{F_{\text{-}}^{\mathit{\sigma}}}$. Furthermore, for nonnegative integers $\sigma$  the representation of $SO(2,1)$ acting in the factor-space $\mathcal{F^{\mathit{\sigma}}}/\mathcal{F_{\text{0}}^{\mathit{\sigma}}}$. These representations are irreducible unitary representations with respect to the scalar product (18)-(19) and is called a discrete series.	
\end{theorem}

\section{SPHERICAL FUNCTIONS OF THE GROUP $SO_0(2,1)$} 
Spherical functions of the group $ SO_0(2,1) $ are one of the important group quantities and in this section we will deal with their direct calculation. A particular case of spherical functions is the zonal functions that are the matrix elements of the ''zero'' column, [6].

The zonal functions of the group $ SO_0(2,1) $ are determined from (8)-(10) as matrix elements of the operator of hyperbolic rotations in the plane $ (k_ {0}, k_ {2}) $ invariant under the subgroup $SO (2)$ of the form of the following integral representation:
\begin{equation}
Z_{\sigma}^{[2,1]}(\alpha)=\frac{1}{2\pi}\intop_{0}^{2\pi}\,\left(\cosh\alpha-\sinh\alpha\cos\phi\right)^{\sigma}d\phi\label{eq:H51}
\end{equation}
We represent the integrand in the form of a Taylor series:
\begin{equation}
\left(\cosh\alpha-\sinh\alpha\cos\phi\right)^{\sigma}=\left(\cosh\alpha\right)^{\sigma}\sum_{n=0}^{\infty}\frac{\left(-\sigma\right)_{n}}{n!}\left(\tanh\alpha\cos\phi\right)^{n}\label{eq:H52}
\end{equation}
Substituting (21) into (20) and integrating term by term for the zonal function of the group $SO_0(2,1)$, we obtain the expression in terms of the Gauss hypergeometric function:
\begin{equation}
Z_{\sigma}^{[2,1]}(\alpha)=\left(\cosh\alpha\right)^{\sigma}\,_{2}F_{1}\left(\frac{1-\sigma}{2},-\frac{\sigma}{2};1;\tanh^{2}\alpha\right).\label{eq:H53}
\end{equation}
It is easy to see that the formula (22) coincides with the expressions for the zonal functions of the group $SO(p,q)$ for $p=2$, [12]. In addition, using (9)-(11) from the integral representation (20) one can obtain:
\[
Z_{-1-\sigma}^{[2,1]}(\alpha)=Z_{\sigma}^{[2,1]}(\alpha).
\]
This is the result of the equivalence of representations $\sigma$ and ($-1-\sigma$).

In a similar way, it is possible to calculate the associated functions of the group $SO_0(2,1)$. For associated functions, we have the following integral representation:
\begin{equation}
P_{\sigma m}^{[2,1]}(\alpha)=\frac{1}{2\pi}\intop_{0}^{2\pi}\,\left(\cosh\alpha-\sinh\alpha\cos\phi\right)^{\sigma}e^{im\varphi}d\phi\label{eq:H54}
\end{equation}
Again, we use the Taylor series (21) and transform the integral (23) to the form:
\[
P_{\sigma m}^{[2,1]}(\alpha)=\frac{\left(\cosh\alpha\right)^{\sigma}}{2\pi}\sum_{n=0}^{\infty}\frac{\left(-\sigma\right)_{n}}{n!}\left(\tanh\alpha\right)^{n}\intop_{0}^{2\pi}\,\left(\cos\phi\right)^{n}e^{im\varphi}d\phi.
\]
The last integral after integration can be reduced to the form:
\begin{equation}
P_{\sigma m}^{[2,1]}(\alpha)=\frac{\left(-\sigma\right)_{m}}{2^{m}m!}\left(\cosh\alpha\right)^{\sigma}\left(\tanh\alpha\right)^{m}\,_{2}F_{1}\left(\frac{m-\sigma}{2},\frac{m-\sigma+1}{2};m+1;\tanh^{2}\alpha\right).\label{eq:H55}
\end{equation}

It is easy to see that the formula (24) coincides with (22) for $m=0$. Moreover, we can verify that the following relation holds:
\begin{equation}
P_{\sigma,-m}^{[2,1]}(\alpha)=(-1)^{m}P_{\sigma m}^{[2,1]}(\alpha).\label{eq:H56}
\end{equation}

The formulas (24) and (25) allow us to obtain associated functions of the group $SO_0(2,1)$ for all integer values of $m$.

\section{THE INDUCED REPRESENTATIONS OF THE GROUP $ISO(2,1)$}

The group $ ISO (2,1) $ is a regular semi-direct product of the group
3-dimensional translations of $T_ {3}$ with the group $SO(2,1)$:
\[
ISO(2,1)=T_{3} \lhd SO(2,1).
\]

 Moreover, the subgroup $ SO (2,1) $ is isomorphic to the automorphism group of the additive subgroup $ T_ {3} $, and $ T_ {3} $ is in turn normal divisor of the group $ ISO (2,1) $. The induced representations of such semi-direct products are constructed by the orbit method, which is a generalization of the Wigner's ''small group'' method in the case of the Poincar\'{e} group, [13].

In this section, instead of using the results of the general theory we shall construct irreducible unitary representations of the group $ ISO (2,1) $ directly.

The element $ g = \left (\vec {a}, \, r \right) $ of the group $ ISO (2,1) $ can be represented in the following splitting form:
\begin{equation}
g=A\left(\vec{a}\right)\Upsilon(r),\label{eq:I12}
\end{equation}
where
\begin{itemize}
\item $ A \left (\vec {a} \right) $ - translation to the vector $ \vec {a} $ in space
$ M_ {3} $ and $ A \left (\vec {a} \right) \in T_ {3} $;
\item $ \Upsilon (r) $ - hyperbolic and orthogonal rotation around the origin of the coordinates in the space $ M_ {3} $ and $ \Upsilon (r) \in SO (2,1) $.
\end{itemize}

The multiplication law for the elements of the group in this notation has the form:
\begin{equation}
\left(\vec{a}_{1},\,r_{1}\right)\times\left(\vec{a}_{2},\,r_{2}\right)=\left(\vec{a}_{1}+r_{1}\vec{a}_{2},\,r_{1}r_{2}\right)\label{eq:I13}
\end{equation}

It is also obvious that, up to isomorphism, we have:
\begin{equation}
A\left(\vec{a}\right)=\left(\vec{a},\,1\right),\qquad\Upsilon(r)= \left(\vec{0},\,r\right).\label{eq:I14}
\end{equation}

It follows from (26) and (28) that restrictions of any irreducible unitary representation of the group $ ISO (2,1) $ on the subgroups $ T_ {3} $ and $ SO (2,1) $ are unitary representations of these subgroups.

The groups $ ISO (2,1) $, induced by representations, will be constructed using the group characters of the subgroup $ T_ {3} $.

As is known from the theory of representations of abelian groups [6], that any irreducible unitary representation of the commutative group is one-dimensional and is represented by an exponent:
\begin{equation}
\chi_{\vec{p}}\left(\vec{a}\right)=e^{i\left(\vec{p}\cdot\vec{a}\right)},\qquad\vec{p}\cdot\vec{a}=p_{0}a_{0}-p_{1}a_{1}-p_{2}a_{2}.\label{eq:I15}
\end{equation}

Here $ \chi _ {\vec {p}} \left (\vec {a} \right) $ is called the character group, and the vector $ \vec {p} $ uniquely characterizes the representation. From (29) it is obvious that the characters of the group form an additive group:
\begin{equation}
\chi_{\vec{p}^{(1)}}\left(\vec{a}\right)\cdot\chi_{\vec{p}^{(2)}}\left(\vec{a}\right)=\chi_{\vec{p}^{(1)}+\vec{p}^{(2)}}\left(\vec{a}\right)
\end{equation}

The choice of a scalar product in the form (29) allows to define an isomorphism of the subgroup $ SO (2,1) $ into the group of automorphisms of group of characters:
\begin{eqnarray}
\left(\Upsilon(r)\chi_{\vec{p}}\right)\left(\vec{a}\right) & = & \chi_{\vec{p}}\left(r\vec{a}\right)=\chi_{r^{-1}\vec{p}}\left(\vec{a}\right).\label{eq:I17}
\end{eqnarray}

Here we used the invariance of the scalar product (29) with respect to the subgroup $ SO (2,1) $. On the other hand, the formula (29) for character allows to determine the action of the translation subgroup $ T_ {3} $:
\begin{equation}
\left(A\left(\vec{b}\right)\chi_{\vec{p}}\right)\left(\vec{a}\right)=\chi_{\vec{p}}\left(\vec{a}+\vec{b}\right)=e^{i\left(\vec{p}\cdot\vec{b}\right)}\chi_{\vec{p}}\left(\vec{a}\right)\label{eq:I18}
\end{equation}

From (31) it is obvious that these characters form the following surface in $ M_ {3} $:
\begin{equation}
p_{0}^{2}-p_{1}^{2}-p_{2}^{2}=m^{2},\label{eq:I19}
\end{equation}

where $m ^ {2}$ is an arbitrary real number characterizing the spectrum irreducible unitary representations of the subgroup $T_ {3}$, which are included in the restriction of an irreducible unitary representation of the group $ISO (2,1)$ to this subgroup. Here, the manifold (33) is an orbit of the subgroup $SO (2,1)$. In other words, the orbit (31) is homogeneous space with respect to the subgroup $SO (2,1)$ and consist of points corresponding to representations subgroups $T_ {3}$ that are included in the restriction of an irreducible unitary representation of the group $ ISO (2,1) $ onto subgroup $T_ {3}$. A quasi-invariant measure on this manifold has the following form:
\begin{equation}
\delta\left(\vec{p}^{\,2}-m^{2}\right)\left(d\vec{p}\right)=\frac{dp_{1}dp_{2}}{2\sqrt{m^{2}+p_{1}^{2}+p_{2}^{2}}}.
\end{equation}

The Wigner's ''small group'' method consists in constructing the representations of the group $ISO (2,1)$ by inducing representations of a stationary subgroup (i.e. ''small group'') of the orbit (33).

Representations of the group $ISO(2,1)$ will be constructed in the space of the group characters of the subgroup $T_ {3}$, i.e. in the space of infinitely differentiable functions with the compact support  $\mathcal{D}_{c}^{\infty}\left (\vec {p};\cdot \right)$, defined on the surface (33) with values in space representations of the stationary subgroup of the orbit (33):\footnote {In quantum mechanics, this construction of representations corresponds to transition from the coordinate representation to the momentum representation.}
\begin{equation}
\left(U(\vec{a},r)f\right)\left(\vec{p}\right)=e^{-i\left(\vec{p}\cdot\vec{a}\right)}\Delta\left(w\left(\vec{p},r\right)\right)f\left(r^{-1}\vec{p}\right),
\end{equation}
where $ w \left (\vec {p}, r \right) $ is an element of a small group, the so-called Wigner rotation, and $ \Delta \left (w \left (\vec {p}, r \right) \right) $ is an irreducible representation of a small group.

To calculate the Wigner rotation, we use the Wigner operator $ h \left (\vec {p}, \mathring {p} \right) $, which takes a fixed point $ \mathring {p} $ to the point $ \vec {p} $:
\begin{equation}
h\left(\vec{p},\mathring{p}\right)\mathring{p}=\vec{p}.
\end{equation}

It is obvious that the Wigner operator $ h \left (\vec {p}, \mathring {p} \right) $, is determined with the accuracy of the element of the stationary subgroup and the choice of one of them, as well as the choice of a fixed point $\mathring {p}$ means choice of representative among the class of equivalent representations of the small group. In the following formulas, the fixed point $ \mathring {p} $ we will not indicate. The Wigner rotation $ w \left (\vec {p}, r \right) $ is determined from the following condition:
\[
r^{-1}h\left(\vec{p}\right)w\left(\vec{p},r\right)=h\left(r^{-1}\vec{p}\right),
\]
i.e.
\begin{equation}
w\left(\vec{p},r\right)=h^{-1}\left(\vec{p}\right)rh\left(r^{-1}\vec{p}\right).\label{eq:I23}
\end{equation}

Otherwise, the operators $ U \left (\cdot, \cdot \right) $ form a representation of the group $ ISO (2,1) $. From the general theory of induced representations of Mackey, it follows that it is irreducible and unitary if the representation $ \Delta \left (\cdot \right) $ of the small group is irreducible and unitary, and the surface (33) is the base of imprimitivity, [13].

Now it is necessary to define transitivity surfaces from (33) and stationary subgroups for different values of $ m ^ {2} $.
\begin{enumerate}
	\item $m^{2}>0.$
	In this case, the surface of transitivity is a two-sheeted hyperboloid. As a fixed point, we choose: $\mathring {p} = (m,0,0)$. Then the stationary subgroup is the rotation group $SO(2)$ on the plane $(p_ {1}, p_ {2})$.\footnote{Obviously, for another choice of a fixed point, we obtain an equivalent stationary subgroup.} On this surface we choose a spherical coordinate system:
	\begin{equation}
	\vec{p}=m(\cosh\alpha,\sinh\alpha\sin\varphi,\sinh\alpha\cos\varphi);\qquad-\infty<\alpha<+\infty,\quad0\leq\varphi<2\pi.\label{eq:I24}
	\end{equation}
	
	The corresponding decomposition of the elements of the subgroup $ SO (2,1) $ and the quasi-invariant measure $ d \mu $ will look like:
	\begin{equation}
	\Upsilon(r)=R(\varphi_{2})h_{02}(\alpha)R(\varphi_{1});\qquad d\mu=\frac{1}{2}\tanh\alpha\,d\alpha\,d\varphi.
	\end{equation}
	
	Here $R (\cdot)$ is the rotation in the plane $(p_ {1}, p_ {2}$), and $ h_ {02} (\cdot) $ -- hyperbolic rotation on the plane $ (p_ {0}, p_ {2} $).
	
	Since the irreducible unitary representations of a small group, i.e. of the rotation group $SO(2)$ on the plane $ (p_ {1}, p_ {2} $) are given an integer $ s $, [6-7], then irreducible unitary representations of group $ ISO(2,1) $ will be given by the numbers $(m,s)$:\footnote {The notation $ (m, s) $ for irreducible unitary representations is chosen from the considerations that in the case of the Poincar\'{e} group,  $ m $ and $ s $ correspond to the mass and spin of elementary particles, respectively.}
	\begin{equation}t
	\left(U^{(m,s)}(\vec{a},r)f\right)\left(\vec{p}\right)=e^{-i\left(\vec{p}\cdot\vec{a}\right)}e^{is\phi}f\left(r^{-1}\vec{p}\right)
	\end{equation}
	Here the Wigner rotation $\phi\left(r,\vec{p}\right)$ is defined
	from the equations (37):
	\begin{equation}
	R(\phi)=h^{-1}\left(\vec{p}\right)rh\left(r^{-1}\vec{p}\right),
	\end{equation}
	where
	\[
	h\left(\vec{p}\right)=R(\varphi)h_{02}(\alpha),
	\]
	is the Wigner's operator.
	
	\item $m^{2}<0.$
	
	In this case, the surface of transitivity is a one-sheeted hyperboloid. As a fixed point, we choose: $ \mathring {p} = (0,0, m) $. Then the stationary subgroup is the hyperbolic rotation group
	on the plane $ (p_ {0}, p_ {1} $). On this surface we choose a hyperbolic coordinate system:
	\begin{equation}
	\vec{p}=m(\sinh\alpha\cosh\beta,\sinh\alpha\sinh\beta,\cosh\alpha),\label{eq:I29}
	\end{equation}
	where $ \alpha, \beta $ are real numbers.
	The corresponding decomposition of the elements of the subgroup $SO (2,1)$ and the quasi-invariant measure $d \mu$	will look like:
	\begin{equation}
		\Upsilon(r)=h_{01}(\beta_{2})h_{02}(\alpha)h_{01}(\beta_{1}),\qquad d\mu=\frac{1}{2}\sinh\alpha d\alpha d\beta.
	\end{equation}
		
	Since the irreducible unitary representations of a small group, i.e. group of hyperbolic rotations $ SH (2) \equiv SO (1,1) $ on the plane $ (p_ {0}, p_ {1} $) are given real number $ s $, [6], then the irreducible unitary representations of the group $ ISO (2,1) $ will be given by the numbers $ (m, s) $: \footnote {In the case of the Poincar\'{e} group this corresponds to elementary systems with continuous spin, [14].}
	\begin{equation}
	\left(U^{(m,s)}(\vec{a},r)f\right)\left(\vec{p}\right)=e^{-i\left(\vec{p}\cdot\vec{a}\right)}e^{is\theta}f\left(r^{-1}\vec{p}\right)
	\end{equation}
	The Wigner rotation $ \theta \left (r, \vec {p} \right) $ is determined from equations (37):
	\begin{equation}
	h_{01}\left(\theta\right)=h^{-1}\left(\vec{p}\right)rh\left(r^{-1}\vec{p}\right),
	\end{equation}
	where
	\[
	h\left(\vec{p}\right)=h_{01}(\beta)h_{02}(\alpha),
	\]
	is the Wigner's operator.
		
	\item $m^{2}=0$, $\vec{p}\neq\vec{0}$.
	
	On this surface we choose a stereographic coordinate system:
		\begin{equation}
		\vec{p}=\tau\left(\frac{1+a^{2}}{2},a,\frac{1-a^{2}}{2}\right);\qquad-\infty<a<+\infty,\quad \tau>0.\label{eq:I36}
		\end{equation}
		
		The corresponding decomposition of the elements of the subgroup $ SO(2,1) $ and the quasi-invariant measure $ d\mu $ will look like:
			\begin{eqnarray}
			\Upsilon(r) & = & B(b)D(\tau)Z(z)=h_{01}(\beta_{2})h_{02}(\alpha)h_{01}(\beta_{1}),\\
			d\mu & = & \tau\left(1-a\right)^{2}d\tau da=e^{2\alpha}\left(1-a\right)^{2}d\alpha da=\frac{1}{2}\sinh\alpha d\alpha d\beta.
			\end{eqnarray}	
	Here
		\begin{equation}
		B(b)=\begin{pmatrix}1+\frac{b^{2}}{2} & a & \frac{b^{2}}{2}\\
			b & 1 & b\\
			-\frac{b^{2}}{2} & -b & 1-\frac{b^{2}}{2}
			\end{pmatrix};\qquad Z(z)=\begin{pmatrix}1+\frac{z^{2}}{2} & z & -\frac{z^{2}}{2}\\
		z & 1 & -z\\
		\frac{z^{2}}{2} & z & 1-\frac{z^{2}}{2}
		\end{pmatrix}.
		\end{equation}
	The matrices $Z(z)$ form the stationary subgroup of fixed point and we choose: $ \mathring {p} = (1/2,0,1 / 2) $.

		The matrices $ B (b) $, as well as the matrices $Z(z)$, form an abelian subgroup.
				
	The Wigner operator:
	\begin{equation}
	h\left(\vec{p}\right)=B(b)D(\tau);\quad D(\tau)=h_{02}(\alpha),\qquad\tau=e^{\alpha};\label{eq:I37}
	\end{equation}
		
	Since irreducible unitary representations of a small group, i.e., one-dimensional commutative group $Z(z)$, are given by a real number $\Lambda$, [6], the irreducible unitary representations of the groups $ ISO (2,1) $ will be determined by the real number $ \lambda $: \footnote {In the case of the Poincar\'{e} group this corresponds to elementary systems with continuous helicity.}
	\begin{equation}
	\left(U^{(0,\lambda)}(\vec{a},r)f\right)\left(\vec{p}\right)=e^{-i\left(\vec{p}\cdot\vec{a}\right)}e^{i\lambda\zeta}f\left(r^{-1}\vec{p}\right)
	\end{equation}
	
	The Wigner rotation $ \zeta \left (r, \vec {p} \right) $ is determined from equations (37):
	\begin{equation}
	h_{02}\left(\zeta\right)=h^{-1}\left(\vec{p}\right)rh\left(r^{-1}\vec{p}\right),
	\end{equation}
	Here the operator 	Wigner $h\left(\vec{p}\right)$ is defined in (50).
	
	\item $m^{2}=0$, $\vec{p}=\vec{0}$.
	
	In this case, the stationary subgroup is the group $SO(2,1)$ and representations of the group $ISO(2,1) $ coincide with the representations of the subgroup $SO(2,1)$, studied in the sections 2-3.
\end{enumerate}

\section{WIGNER's COEFFICIENTS OF THE GROUP $SO_{0}(2,1)$}
In this section we study invariant three-linear functionals in the space
$\mathcal{F}^{\sigma_{1}}\otimes\mathcal{F}^{\sigma_{2}}\otimes\mathcal{F}^{\sigma_{3}}$.
According to the kernel theorem, the invariant three-linear forms will have
the following form, [8]:
\begin{equation}
(f^{(1)},f^{(2)},f^{(3)})=\frac{1}{\left(2\pi\right)^{3}}\intop_{0}^{2\pi}\intop_{0}^{2\pi}\intop_{0}^{2\pi}K^{(3)}(\phi_{1},\phi_{2},\phi_{3})f^{(1)}(\phi_{1})f^{(2)}(\phi_{2})f^{(3)}(\phi_{3})d\phi_{1}d\phi_{2}d\phi_{3}.\label{eq:51}
\end{equation}

Here $K^{(3)}(\phi_{1},\phi_{2},\phi_{3})$ -- is a generalized function over the space $\mathcal{F}^{\sigma_{1}}\otimes\mathcal{F}^{\sigma_{2}}\otimes\mathcal{F}^{\sigma_{3}}$.
It follows from the invariance condition and also from (7)-(8) that $K^{(3)}(\phi_{1},\phi_{2},\phi_{3})$ is a function depending on differences of variables $\phi_{1}-\phi_{2},\,\phi_{2}-\phi_{3},\,\phi_{3}-\phi_{1}$ and satisfying the following equation:
\begin{equation}
	K^{(3)}(\phi_{1\alpha},\phi_{2\alpha},\phi_{3\alpha})=\omega_{\alpha}^{\sigma_{1}+1}\left(\phi_{1}\right)\omega_{\alpha}^{\sigma_{2}+1}\left(\phi_{2}\right)\omega_{\alpha}^{\sigma_{3}+1}\left(\phi_{3}\right)K^{(3)}(\phi_{1},\phi_{2},\phi_{3}),\label{eq:52}
\end{equation}
where $\omega_{\alpha}(\cdot)$ is defined in (9).

It is possible to show that the functional equation (54) has a solution with the accuracy of constant:
\begin{equation}
K^{(3)}(\phi_{1},\phi_{2},\phi_{3})=c\,\frac{\left[1-\cos\left(\phi_{1}-\phi_{2}\right)\right]^{\beta_{3}}}{\Gamma\left(\beta_{3}+1/2\right)}\,\frac{\left[1-\cos\left(\phi_{2}-\phi_{3}\right)\right]^{\beta_{1}}}{\Gamma\left(\beta_{1}+1/2\right)}\,\frac{\left[1-\cos\left(\phi_{3}-\phi_{1}\right)\right]^{\beta_{2}}}{\Gamma\left(\beta_{2}+1/2\right)}.
\end{equation}
Here
\begin{equation}
\beta_{1}=\frac{\sigma_{1}-\sigma_{2}-\sigma_{3}-1}{2},\quad \beta_{2}= \frac{\sigma_{2}-\sigma_{3}-\sigma_{1}-1}{2},\quad \beta_{3}=\frac{\sigma_{3}-\sigma_{2}-\sigma_{1}-1}{2}.
\end{equation}

Wigner coefficients are defined as the value of the functional (53) on elements of a canonical basis:
\begin{eqnarray}
\begin{pmatrix}\sigma_{1} & \sigma_{2} & \sigma_{3}\\
m_{1} & m_{2} & m_{3}
\end{pmatrix} & = & \frac{c}{\left(2\pi\right)^{3}}\intop_{0}^{2\pi}\intop_{0}^{2\pi}\intop_{0}^{2\pi}\frac{\left[1-\cos\left(\phi_{1}-\phi_{2}\right)\right]^{\beta_{3}}}{\Gamma\left(\beta_{3}+1/2\right)}\,\frac{\left[1-\cos\left(\phi_{2}-\phi_{3}\right)\right]^{\beta_{1}}}{\Gamma\left(\beta_{1}+1/2\right)}\times\nonumber \\
& \times & \frac{\left[1-\cos\left(\phi_{3}-\phi_{1}\right)\right]^{\beta_{2}}}{\Gamma\left(\beta_{2}+1/2\right)}e^{im_{1}\phi_{1}}e^{im_{2}\phi_{2}}e^{im_{3}\phi_{3}}d\phi_{1}d\phi_{2}d\phi_{3},
\end{eqnarray}
where $\beta_{1},\beta_{2},\beta_{3}$ are defined in (56). 

Since the exponent is orthonormal it follows that the last expression is not zero only under condition:
\begin{equation}
m_{1}+m_{2}+m_{3}=0,
\end{equation}
and after necessary integrations and transformations this expression
is reduced to:
\begin{equation}
\begin{pmatrix}\sigma_{1} & \sigma_{2} & \sigma_{3}\\
m_{1} & m_{2} & m_{3}
\end{pmatrix}= c\frac{2^{-\left(\sigma_{1}+\sigma_{2}+\sigma_{3}+3\right)/2}}{\pi^{3/2}}\times\nonumber 
\end{equation}
\begin{equation}
\times  \sum_{n=-\infty}^{+\infty}\frac{(-\beta_{3})_{n-m_{1}}(-\beta_{1})_{n+m_{3}}(-\beta_{2})_{n}}{\Gamma\left(n-m_{1}+\beta_{3}+1\right)\Gamma\left(n+m_{3}+\beta_{1}+1\right)\Gamma\left(n+\beta_{2}+1\right)}.\label{eq:57}
\end{equation}
Here we have taken into account that
\[
\beta_{1}+\beta_{2}+\beta_{3}=-\frac{\sigma_{1}+\sigma_{2}+\sigma_{3}+3}{2}.
\]

The formula (\ref{eq:57}) can be reduced to the following form:
\begin{eqnarray}
\begin{pmatrix}\sigma_{1} & \sigma_{2} & \sigma_{3}\\
m_{1} & m_{2} & m_{3}
\end{pmatrix} & = & \frac{c(-1)^{m_{1}}\left[2^{\sigma_{1}+\sigma_{2}+\sigma_{3}+3}\pi\right]^{-1/2}}{\Gamma\left(\beta_{2}+1\right)\Gamma\left(-m_{1}+\beta_{3}+1\right)\Gamma\left(m_{3}+\beta_{1}+1\right)}\frac{\left(-\beta_{1}\right)_{m_{3}}}{\left(1+\beta_{3}\right)_{m_{1}}}\times\nonumber \\
& \times & \,_{3}H_{3}\begin{bmatrix}-\beta_{3}-m_{1}, & -\beta_{1}+m_{3}, & -\beta_{2};\\
&  &  & 1\\
-m_{1}+\beta_{3}+1, & m_{3}+\beta_{1}+1, & \beta_{2}+1;
\end{bmatrix},\label{eq:58}
\end{eqnarray}
where $\,_{3}H_{3}$ -- is the bilateral series [11].

The arbitrary constant $c$ is defined so that the following condition is satisfied:
\begin{equation}
\begin{pmatrix}\sigma_{1} & \sigma_{2} & \sigma_{3}\\
0 & 0 & 0
\end{pmatrix}=1.\label{eq:510}
\end{equation}

Then from (\ref{eq:58})-(\ref{eq:510}) for the constant $c$ we get:
\begin{equation}
c=\sqrt{2^{\sigma_{1}+\sigma_{2}+\sigma_{3}}\pi}\,\frac{\Gamma\left(\beta_{1}+\beta_{2}+1\right)\Gamma\left(\beta_{1}+\beta_{3}+1\right)\Gamma\left(\beta_{2}+\beta_{3}+1\right)}{\Gamma\left(\beta_{1}+\beta_{2}+\beta_{3}+1\right)}.
\end{equation}

Finally, taking into account (60) we get the following expression for Wigner coefficients:
\begin{eqnarray}
\begin{pmatrix}\sigma_{1} & \sigma_{2} & \sigma_{3}\\
m_{1} & m_{2} & m_{3}
\end{pmatrix} & = & \frac{(-1)^{m_{1}}\delta_{m_{1}+m_{2}+m_{3},0}}{\Gamma\left(\beta_{2}+1\right)\Gamma\left(-m_{1}+\beta_{3}+1\right)\Gamma\left(m_{3}+\beta_{1}+1\right)}\times\nonumber \\
& \times & \frac{\Gamma\left(\beta_{1}+\beta_{2}+\beta_{3}+1\right)}{\Gamma\left(\beta_{1}+\beta_{2}+1\right)\Gamma\left(\beta_{1}+\beta_{3}+1\right)\Gamma\left(\beta_{2}+\beta_{3}+1\right)}\times\nonumber
\end{eqnarray}
\begin{equation}
\times\frac{\left(-\beta_{1}\right)_{m_{3}}}{\left(1+\beta_{3}\right)_{m_{1}}}\,_{3}H_{3}\begin{bmatrix}-\beta_{3}-m_{1}, & -\beta_{1}+m_{3}, & -\beta_{2};\\
&  &  & 1\\
-m_{1}+\beta_{3}+1, & m_{3}+\beta_{1}+1, & \beta_{2}+1;
\end{bmatrix}.\
\end{equation}
Here $\beta_{1},\beta_{2},\beta_{3}$ are defined in (56).

From the invariance of three-linear form (53) implies the covariance property of the Wigner coefficients (57):
 {\begin{equation}
 	\begin{pmatrix}\sigma_{1} & \sigma_{2} & \sigma_{3}\\
 	m_{1} & m_{2} & m_{3}
 	\end{pmatrix}=\sum_{m_{1}^{'}m_{2}^{'}m_{3}^{'}}\,\begin{pmatrix}\sigma_{1} & \sigma_{2} & \sigma_{3}\\
 	m_{1}^{'} & m_{2}^{'} & m_{3}^{'}
 	\end{pmatrix}\prod_{i=1}^{3}t_{m_{i}^{'}m_{i}}^{\sigma_{i}}\left(g\right),\qquad\forall g\in SO_0(2,1).
 	\end{equation}
 Here, $t_{m^{'},m}^{\sigma}(g)$ are the matrix elements of the unitary irreducible representations of the group $SO_0(2,1)$ in the canonical basis (6).
 
 The covariance property (64) plays a fundamental role when constructing the multi-linear invariants of the representation of the group $SO_0 (2,1)$.
 
\section*{CONCLUSION}
It is important to compare results obtained for the group $SO(2,1)$ with the theory of representations of the Lorentz group $SO(3,1)$, [5-7, 13]. It can be seen from an analysis given in Section 3, representations of the group $SO(2,1)$, despite the dimension reduction, have more complex structure. 

In conclusion, I would like to thank Prof. A.Baghirov, Prof. N.Atakishiyev and Prof. E.Veliev  for their attention to this work and discussion of results.

\section*{\textbf{Appendix.} FOURIER EXPANSION OF THE FUNCTION $\left(1-\cos\psi\right)^{\lambda}$.}
We will formulate the Fourier-expansion of the function $\left(1-\cos\psi\right)^{\lambda}$ where $\lambda\in\mathbb{C}$ is a complex number:
\begin{eqnarray}
\left(1-\cos\psi\right)^{\lambda} & = & \sum_{m=-\infty}^{+\infty}\,a_{m}e^{im\psi},\label{eq:31}\\
a_{m} & = & \frac{1}{2\pi}\intop_{0}^{2\pi}\left(1-\cos\psi\right)^{\lambda}e^{-im\psi}d\psi.
\end{eqnarray}

Applying Taylor series for the function $\left(1-\cos\psi\right)^{\lambda}$:
\[
\left(1-\cos\psi\right)^{\lambda}=\sum_{n=0}^{\infty}\frac{\left(-\lambda\right)_{n}}{n!}\left(\cos\psi\right)^{n}
\]
and using the Gauss formula for a hypergeometric function of the unit variable [6, 9, 11], one can get the following expression for coefficients $a_{m}$$\left(-\lambda\right)_{n}$:
\begin{equation}
a_{m}=\frac{2^{\lambda}\Gamma(\lambda+1/2)}{\sqrt{\pi}\Gamma(m+\lambda+1)}(-\lambda)_{m}.
\end{equation}

In these formulas we used the notation $\left(-\lambda\right)_{n}$ for the Pochhammer symbol [9, 11]. Thus the expansion (65) has the form:
\begin{equation}
\left(1-\cos\psi\right)^{\lambda}=\frac{2^{\lambda}\Gamma(\lambda+1/2)}{\sqrt{\pi}}\sum_{m=-\infty}^{+\infty}\frac{(-\lambda)_{m}}{\Gamma(m+\lambda+1)}e^{im\psi}.
\end{equation}

It can be seen from here that the integral (66) converges when $\mathrm{Re}\,\lambda>-1/2$. This integral can be considered in the sense of regularized values [10] when $\mathrm{Re}\,\lambda<-1/2$.




\bigskip

\begin{thebibliography}{10}

\bibitem{} A.D. Alhaidari, arXiv:math-ph/0112004, 2002, p.14

\bibitem{} Gaetano Vilasi and Patrizia Vitale, arXiv:gr-qc/0202018v.1, 2002, p.10

\bibitem{} Exact solutions of the Einsteins field equations, ed. E.Schmutzer, Berlin, 1980.

\bibitem{} Blagoje Oblak, BMS Particles in Three Dimensions, Springer, 2017, 461 p. 

\bibitem{} Gelfand I.M., Graev  M.I.. Vilenkin  H.Y. Integral geometry and associated questions of the theory of representation, M.: FM 1962, 656 p. (In Russian).

\bibitem{} Vilenkin N.Y,  Special functions and theory of groups representation, M.: Nauka, 1965, 588 p. 

\bibitem{} Zhelobenko D.P., Stern A.I., Representations of Lie Groups, (Moscow:Nauka), 1983. (in Russian)

\bibitem{} Treves F.,Topological Vector Spaces, Distributions and Kernels, (1967), Purdue University, Indiana.

\bibitem{} Bateman, G.,Erd\'{e}lyi A., Higher transcendental functions, M.: Nauka,1973, v.1, 294 p. (In Russian)..

\bibitem{} Gelfand I.M.,Shilov G.E. Generalization of the function and operations on them (Generalized Functions, vol. 1, 2nd edition) - M.: Physmatlit, 1959, 471 p. (In Russian).

\bibitem{} Slater L.J. Generalized hypergeometric functions – Cambridge, 1966, 273 p.

\bibitem{} Rajabov B.A.,Tr.J. of App.Eng.Math., N.2, (2017)

\bibitem{} A.O. Barut, R. Raczka, (1977), Theory of Group Representations and Applications, Warszawa.

\bibitem{} Xavier Bekaert, Evgeny Skvortsov, arXiv:1708.01030v2 [hep-th], 2017, p.33.

  
\end{thebibliography}
\end{document}